\documentstyle[preprint,aps,prl]{revtex}
\begin{document}
\newcommand{\be}{\begin{equation}}
\newcommand{\ee}{\end{equation}}
\bibliographystyle{unsrt}
\tightenlines
\title{Zeroth and Second Laws of Thermodynamics
Simultaneously Questioned in the Quantum Microworld}
\author{V. \v{C}\'{a}pek}
\address{Institute of Physics of Charles University, Faculty of Mathematics
and Physics,\\ Ke Karlovu 5, 121 16 Prague 2, Czech republic
\\(Tel. (00-420-2)2191-1330, Fax (00-420-2)2492-2797,  \\E-mail
capek@karlov.mff.cuni.cz),}
\date{
January 19, 2001
}
\maketitle
\begin{abstract}
Several models of quantum open systems are known at present to violate,
according to principles of the standard quantum theory of open systems,
the second law of thermodynamics. Here, a new and rather trivial model of
another type is suggested describing mechanism that violates, according to
the same principles, the zeroth and the second laws of thermodynamics
simultaneously. Up to a technically minor modification, the model
resembles some models already known, solved by standard means, and
properly understood. Universal validity of two basic principles of
thermodynamics in {\em strictly quantum situations} is thus simultaneously
called in question.
\end{abstract}

\pacs{~}

\narrowtext
\newpage

\section{Introduction}

Phenomenological arguments against general validity of standard statistical
thermodynamics and call for inclusion of cooperative, selforganization and
similar complicated phenomena even in absence of, e.g., external flows and
far from equilibrium exist already for a long time. Since early nineties,
one can find them especially in, e.g., theory of electron-transfer chemical
reactions where, in connection with their phenomenological non-linear
description, inclusion of such effects seems to be indispensable
\cite{TriPoh}. Necessity of their inclusion follows also from detailed
analysis of what is known in molecular biology about how individual
molecules (molecular machines) work in living organisms \cite{Good}.
Selforganization is usually believed to be a domain of nonlinear theories.
In 1996, the first hamiltonian linear quantum model was suggested that can
lead to a selforganized state upon thermalization in a bath even when no
external flows exist and, simultaneously, this state is
energetically disadvantageous \cite{Cap1,Cap2}. Recent analysis has
revealed that the former non-linear phenomenological and the latter linear
first-principle type of reasoning strive in the same direction and can
be easily united \cite{CapTri}. The reader is referred to
\cite{Cap3,CapBok,CapMan} or also to \cite{CapTri} for previous models
(extending also reasoning of \cite{Cap1,Cap2}) where the cooperative and
selforganizational tendencies in such models can be shown to
provide a basis and unique possibility of violating even the second law of
thermodynamics.

Some of such models allow a mathematically rigorous treatment throughout
all the calculations (\cite{Cap3} is perhaps the first one of them).
However, for technical reasons, just one-step processes have been treated
so far. Only very recently, first in 1998, the first rigorously solvable
quantum model working cyclically as a {\it perpetuum mobile} of the second
kind (i.e. converting heat from a single bath into, this time, a usable
work without compensation) and violating thus, for the first time
explicitly, the Thomson formulation of the second law of thermodynamics
\cite{LieYng} was reported \cite{MECO23,CapBok2,CapBok3}. In 1999 and 2000,
also other groups arrived, independently and for other situations, at the
same conclusion challenging universal validity of the second law
\cite{Niku,AllNie,AllNie2}. From them, in particular paper \cite{AllNie} by
Allahverdyan and Nieuwenhuizen inspired a public response \cite{Weis}.
Completely different mechanism (connected with dynamically maintained
steady-state pressure gradients in rarified gasses) potentially allowing
violation of the second law was recently suggested by Sheehan - compare, in
connection with previous paper \cite{Shee1}, the discussion in
\cite{Dunc,Shee2}. Another and even rather positively experimentally
tested system was suggested by Sheehan already in 1994
\cite{Shee3,Jone,Shee4}. Because of complicated nature of the problem as
well as owing to 150 years of traditionally presumed universal validity of
thermodynamic principles, it is likely that irrespective of final result
of the above counter-examples, the problem of potential violation of
thermodynamic principles will remain topical for many years to come. A
comment is only worth mentioning here that the above models of the present
group always work from strong correlations (entanglement) among particles
and/or competing and mutually interfering reaction (transfer) channels.
Thus, like in \cite{Niku,AllNie}, the mechanisms discussed are appreciably
different from those based on, e.g., the Feynman ratchet and pawl systems
\cite{Feyn,Muss}. It is most likely that such ratchet-like systems (that
would transform the thermal noise into one-directional linear or rotational
motion) rather fail in experiment once they are devised to violate the
second law \cite{Muss}, irrespective of previous opposite expectations but
fully in accordance with the Feynman original analysis \cite{Feyn}.

The above models of the present group (usually described as isothermal
Maxwell demon models because of dynamic opening and closing `gates', i.e.
reaction channels) are in fact so far sufficiently complicated. Moreover,
their classical counterparts do not work. All this is why they can and
really do, at the first inspection, naturally induce fully comprehensible
prejudice, mistrust, or misunderstandings. To this mistrust, also the fact
contributes that the principles on which their activity relies remind of
the original Maxwell demon \cite{Maxw} (opening and closing a
gate after checking state or performance of previous steps and thus
deciding about next elementary steps in, e.g., particle transfer). The idea
of Maxwell demon was, however, often and in detail analyzed during the
last 130 years \cite{Szil,LefRex}. Result of the analysis was
usually negative but it well applies to just classical models. In this
connection, it is worth mentioning that all models
\cite{Cap1,Cap2,CapTri,Cap3,CapBok,CapMan,MECO23,CapBok2,CapBok3,Niku,AllNie}
really cease to work in the classical (high-temperature) limit. On the
other hand, this analysis does not regard purely quantum models working on,
e.g., principles known from nature, in particular from the contemporary
molecular biology (interplay between particle transfer and accompanying
topological reconstruction of the particle surroundings \cite{Good}) built
in \cite{Cap1,Cap2,CapTri,Cap3,CapBok,CapMan,MECO23,CapBok2,CapBok3}.

The situation seems to be even more serious as the quantum model of the
isothermal Maxwell demon of \cite{Cap1,Cap2} (model of uni-directional
isothermal particle transfer even against potential forces) allows a simple
generalization to a greater set of sites available to a greater number
of particles. This analysis then implies that in a stationary state
(equilibrium one in the sense of thermodynamics), chemical potentials of
one sort of particles in two subsystems interconnected by sophisticated
(e.g. molecular) bridges could become even different \cite{CaSimem}. This
questions universal validity of another basic principle of the statistical
thermodynamics \cite{LanLif}. In order to make the situation simpler, we
have here rebuilt the model of \cite{Cap1,Cap2} to describe transfer
of excitons (i.e. excitation energy) and analyzed principles of its
work in connection with those of the original model. It appeared that the
rebuilt model could be appreciably simplified to such an extent that it
becomes fully independent. No possibility is seen to simplify it further
and to preserve, simultaneously, the unusual phenomena investigated. The
above elementary steps of `checking performance of previous steps' and
`opening or closing the gate according to the result of the check', so
typical of the Maxwell demon - like models, completely disappeared. What,
on the other hand, remained is the existence of quantum interference of
different reaction (transfer etc) channels. Except for a small (rather
physical than technical) modification connected with presumed initial
conditions and existence of two thermodynamic baths, the form of the model
is fully standard. Without this modification, its solution is known, has
been obtained many times and in many different ways, and is correspondingly
believed to be well understood. Simplicity of the model and many times
verified applicability of the really standard technical `weaponry' applied
to it is what may then, hopefully, change the so far reserved attitude of
general public to the above provoking ideas questioning, on grounds of the
quantum theory of open systems, universal validity of principles of the
statistical thermodynamics. These questioned principles include now, in the
light of the present results, not only the second but, as argued below,
also the zeroth law of thermodynamics.

\section{Model}

System of a few levels interacting with a thermodynamic bath is a standard
quantum problem. We shall use it also here. Specifically, here, we assume
three levels, one ground and two excited ones, and refer to the excited
levels as those with a Frenkel exciton placed either on site 1 or site 2.
These sites might be, e.g., two different molecules or local centres in,
possibly, two different but adjacent solids representing two different
electron subsystems. As it is easy to verify, it is for our problem here
irrelevant whether we include or ignore the fourth level corresponding to
both molecules (molecular systems, local centres) excited. For technical
simplicity, we choose the latter alternative. Corrections owing to the
(here ignored) two-exciton states are, in, e.g., the excited state
occupation probabilities and for small $1\leftrightarrow2$ exciton
transfer rates, $\propto\exp\{-\beta(\epsilon_1+\epsilon_2)\}$
where $\epsilon_j$, $j=1,2$ are the local exciton energies. Thus, they
can be easily identified.

The exciton residing possibly at sites 1 or 2 (if not lacking at all for
some time owing to finite life-time effects admitted here) can in
principle be transferred between the sites (subsystems) either coherently
or incoherently. Here, we choose the first alternative, designating the
hopping (resonance or transfer) integral as $J$. Finally, we complete the
model by adding a bath interacting with the system. The tricky feature
whose real sense will be seen only below is that we ascribe to each
subsystem (designated as I with site 1 and II with site 2) its own
thermodynamic bath represented by harmonic oscillators (phonons). Because
we assume that initially, both the baths have a canonical distribution with
possibly equal temperatures, this step is in such a case isomorphic to
assuming that the exciton at site 1 or 2 interacts respectively with, e.g.,
just even or odd modes of a single bath. Technically,
the next step is not relevant but physically, it is important to assume
that these two baths are in a way separated as we shall discuss the energy
(heat) flow from bath I to II and vice versa. Adding that the
exciton-phonon coupling (with coupling constants $g_{\kappa}$ and
$G_{\kappa}$) leading to the above exciton finite life-time effects (i.e.
non-preserving the number of excitons) is assumed linear in the phonon
creation ($b_{j\kappa}^{\dagger}$) and annihilation ($b_{j\kappa}$, $j=1$
and $2$) operators, one can directly write the Hamiltonian as
\[ H=H_I+H_{II}+J(a_1^{\dagger}a_2+a_2^{\dagger}a_1),\]
\[ H_I=\epsilon_1a_1^{\dagger}a_1+\sum_{\kappa}\hbar\omega_{\kappa}
b_{1\kappa}^{\dagger}b_{1\kappa}+\frac{1}{\sqrt{N}}\sum_{\kappa}
g_{\kappa}\hbar\omega_{\kappa}(a_1+a_1^{\dagger})(b_{1\kappa}+
b_{1\kappa}^{\dagger}), \]
\be H_{II}=\epsilon_2a_2^{\dagger}a_2+\sum_{\kappa}\hbar\omega_{\kappa}
b_{2\kappa}^{\dagger}b_{2\kappa}+\frac{1}{\sqrt{N}}\sum_{\kappa}
G_{\kappa}\hbar\omega_{\kappa}(a_2+a_2^{\dagger})(b_{2\kappa}+
b_{2\kappa}^{\dagger}). \label{Ham} \ee
The exciton creation (annihilation) operators are
assumed to fulfil the Pauli relations
\[ \{a_1,a_1^{\dagger}\}=\{a_2,a_2^{\dagger}\}=1, [a_1,a_2^{\dagger}]=0,\]
\be \{a_1,a_1\}=\{a_2,a_2\}=[a_1,a_2]=0\;{\rm etc.} \label{Paucom} \ee
Here, $\{\ldots,\ldots\}$ and $[\ldots,\ldots]$ are the usual anti- and
commutators. The phonon frequencies $\omega_{\kappa}$ are, for simplicity,
assumed the same for both the baths. Finally, $\epsilon_j$, $j=1,\;2$ are
the exciton energies while $N$ is the number of the phonon modes (finite
before taking the baths thermodynamic limit) in each bath separately.
Technically, the existence of two separated and uncorrelated baths, each of
them interacting with just one exciton level, makes also the form of our
matrices below simpler.

We now want to proceed by writing down a closed set of equations for the
exciton density matrix only, projecting off the information about baths.
There are several well known ways leading finally to the same result. In
order to be specific, we choose time-convolutionless Generalized master
equations \cite{Fuli,FulKra,SHTS1,SHTS2,Gzyl} with the Argyres-Kelley
projector \cite{ArgKel,Peie}. Practical application of the resulting
equations then requires approximations (e.g. expansions) upon calculation
of coefficients in the equations that can be avoided by application of
scaling arguments.
The method is well developed and standard now. Describing the formal
apparatus, we must be, however, more specific as we are now going to
deviate from a standard weak-coupling approach. This is necessary
here because of necessity to describe properly interplay between/among
several competing processes.

The weak coupling theory was made mathematically perfect by Davies who
properly implied the scaling idea of van Hove \cite{Davi1,Davi2}.
It is funny to realize that, at least for systems of finite number of
levels, the mathematical prescription how to calculate properly the
weak-coupling dynamics as provided by the Davies theory is not unique.
For that, compare Theorem 1.4 of \cite{Davi2} which establishes full
uniqueness just in the full Van Hove limit [see (\ref{VanHove}) below] but
not for arbitrarily weak though finite coupling strengths as it
corresponds to reality. Irrespective of it, because of its mathematically
rigorous form, this theory (really rigorous in its region of validity) so
influenced theoreticians that many of them now consider the weak
coupling language as universal. This is, of course, unintentional negative
consequence of the mathematical precision of the Davies theory. So, it is
also (and even more than above) funny to observe that this theory fails
to describe, e.g., the physical regime we are here interested in. This is
owing to the overestimation, in the Davies (or, more generally, weak
coupling) theory, of the role of in-phasing (that is, in our case, owing
to, e.g., the coherent exciton transfer $1\leftrightarrow2$) as compared
to the standard dephasing.
\footnote{This overestimation is not owing to any formal error, it is
because of the very form of the presumed Van Hove scaling. In Nature,
there is, e.g., no possibility to scale coupling constants as they are
real constants, not variables.} The latter
dephasing process is, in our case, simply due to the exciton (electron)
coupling to the bath. Let us be even more specific:

The standard weak-coupling theory is based on the notion of a small
pa\-ram\-e\-ter, say $g$, of the coupling of the system to the bath. This
means that $g$ would be a joint small parameter of the last terms on the
right hand sides of $H_I$ and $H_{II}$ in (\ref{Ham}) only. Then a new
unit of time, say $\tau=t/t'$ is chosen and we work, instead of the true
physical time $t$, in terms of the dimensionless time $t'$.
Avoiding here technical details how to avoid Poincar\'{e} cycles by
performing first the thermodynamic limit of the bath \cite{FicSau}, the
result is that finally, the proper weak-coupling equations for the
exciton density matrix are obtained (otherwise as below) by taking the
combined limiting Van Hove procedure
\be g\rightarrow0,\;\tau\rightarrow+\infty,\; g^2\tau={\rm const}.
\label{VanHove} \ee
Omitting at this moment also less important mathematical details (see
\cite{Davi2,Davi3}), the statement is that the resulting
equations then describe properly the time development of the system, in
terms of the new time $t'$, to its canonical state corresponding to the
(initial) temperature of the bath. Because of the limit $g\rightarrow0$
incorporated into (\ref{VanHove}) and because the coherent (transfer or
hopping) integrals responsible for the in-phasing are kept finite during
this limit, the role of the dephasing processes in competing the in-phasing
ones is fully suppressed. This is why the the corresponding asymptotic,
i.e. canonical density matrix is then diagonal in the basis of the
eigenstates of the Hamiltonian of the system $H_S$ alone. In other words,
this is why the relaxation goes to eigenstates of $H_S$.

In order to incorporate the above competition between dephasing and
in-phasing processes, we shall proceed in almost the same way except for
one point: We take $g$ as a small parameter of not only the system coupling
to the bath. We also assume that $J\propto g^2$. The resulting limiting
procedure is thus not that of the weak coupling but that of, rather, slow
transfer processes. Concerning the relative strength of the
$1\leftrightarrow2$ transfer and relaxation owing to the coupling to the
bath, it might be, in such a scheme and for a general situation, still
arbitrary. Here, we shall, however, assume that the latter coupling is
rather intermediate or even strong as compared to the coherent
$1\leftrightarrow2$ transfer of the exciton in the sense that the
dephasing is either comparable to, or even dominates over in-phasing.
(Concerning the words `...dominates over..' remember, that in the
situation of the weak coupling case, the in-phasing is, in the limiting
sense of (\ref{VanHove}), {\em infinitely} stronger than the dephasing.
Here, on the contrary, we admit in our case of the intermediate or strong
coupling that the dephasing could be even dominating over the in-phasing
but, figuratively speaking, their ratio remains always {\em finite} though
perhaps arbitrarily large.) This
causes remarkable differences in structure of the equations we aim at as
well as in the physical conclusions. In particular, we then get that the
relaxation does not go (in the sense of diagonalizing the asymptotic
density matrix) to the eigenstates of the Hamiltonian of the system. To
what state (i.e. to what exciton density matrix) the relaxation then
goes we shall see later.

Technically, the method of deriving the closed set of equations for the
exciton density matrix proceeds in the following steps: \begin{itemize}
\item We introduce the density matrix of the `system+bath(s)' complex
in the `interaction' picture as
\be \tilde{\rho}(t)=\exp\{i{\cal L}_0\cdot t\} \rho_{S+B}(t).
\label{intpic} \ee
Here $\rho_{S+B}(t)$ is the density matrix of the system and the bath in
the Schr\"{o}\-din\-ger picture and the Liouvillean ${\cal L}_0\ldots=
[H_0,\ldots]/\hbar$ where, however,  $H_0$ $=\sum_{j=1}^2[\epsilon_j
a_j^{\dagger}a_j+\sum_{\kappa}\hbar\omega_{\kappa}b_{j\kappa}^{\dagger}
b_{j\kappa}]$. In
other words, the hopping term $J(a_1^{\dagger}a_2+a_2^{\dagger}a_1)$ is now
{\em not} included in $H_0$. This is unlike the scaling inherent to
the weak-coupling case (\ref{VanHove}).
\item We apply, e.g., the Fuli\'{n}ski and
Kramarczyk identity \cite{Fuli,FulKra}, or its more famous form by
Shibata, Hashitsume, Takahashi, and Shingu \cite{SHTS1,SHTS2}
\[ \frac{d}{dt}{\cal P}\tilde{\rho}(t)=-i{\cal PL}(t)[1+i\int_0^t
\exp_{\leftarrow}\{-i\int_{\tau_1}^t(1-{\cal P}){\cal L}(\tau_2)\,d\tau_2
\}(1-{\cal P}){\cal L}(\tau_1)\] \[\cdot\exp_{\rightarrow}\{i\int_{\tau1}^t
{\cal L}(\tau_2)\,d\tau_2\}\,d\tau_1]^{-1}\] \be\cdot
[\exp_{\leftarrow}\{-i\int_0^t(1-{\cal P})\}
{\cal L}(\tau)\,d\tau\}(1-{\cal P})\rho(0)+{\cal P}\tilde{\rho}(t)].
\label{SHTS} \ee
(For equivalence of (\ref{SHTS}) with \cite{Fuli,FulKra} see \cite{Gzyl}.)
Here ${\cal P}$ is the so called Argyres-Kelley projector
\cite{ArgKel,Peie}
\be {\cal P}\ldots=\rho^B\,Tr_B(\ldots), \label{ArgKel} \ee
\be {\cal L}(t)\ldots=\exp\{i{\cal L}_0\cdot
t\}\frac{1}{\hbar}[\tilde{H},\ldots]\exp\{-i{\cal L}\cdot t\},
\label{IntLi} \ee and
\[\tilde{H}=J(a_1^{\dagger}a_2+a_2^{\dagger}a_1)\]
\be+\frac{1}{\sqrt{N}}\sum_{\kappa}
\hbar\omega_{\kappa}[g_{\kappa}(a_1+a_1^{\dagger})(b_{1\kappa}+
b_{1\kappa}^{\dagger})+G_{\kappa}(a_2+a_2^{\dagger})(b_{2\kappa}+
b_{2\kappa}^{\dagger})]. \label{H_1} \ee
\item Assume, for simplicity, initially factorizable density matrix of the
exciton and bath and identify $\rho^B$ with the initial density matrix of
the bath.
\item Perform the above scaling (including transition to new time $t'$ -
we shall, however, continue writing $t$) with also $J\propto g^2$,
\item Return back from the
above `interaction' picture to the Schr\"{o}dinger picture. \end{itemize}
(Details of this procedure may be found, for similar models, elsewhere
- see, e.g., \cite{CapBar} for a method fully following the Davies approach
\cite{Davi2,Davi3}.) Then, after some straightforward algebra, the
required set of equations for the density matrix of the exciton only
(designated as $\rho(t)$), that is exact in the sense of the above scaling
procedure, reads
\[ \frac{d}{dt}\left(\begin{array}{c}\rho_{00}(t) \\ \rho_{11}(t) \\
\rho_{22}(t) \\ \rho_{12}(t) \\ \rho_{21}(t) \end{array}\right)=
\left(\begin{array}{cc} {\cal A} & {\cal B}\\ {\cal C} & {\cal D}
\end{array}\right)
\cdot\left(\begin{array}{c} \rho_{00}(t) \\ \rho_{11}(t) \\
\rho_{22}(t) \\ \rho_{12}(t) \\ \rho_{21}(t) \end{array}\right). \]
\be \label{set1} \ee
Here the blocks
\[ {\cal A}=\left(\begin{array}{ccc} -\gamma_{\uparrow}-\Gamma_{\uparrow}
& \gamma_{\downarrow} & \Gamma_{\downarrow} \\
\gamma_{\uparrow} & -\gamma_{\downarrow} & 0 \\
\Gamma_{\uparrow} & 0 & -\Gamma_{\downarrow} \end{array}\right),\;
{\cal B}=\left(\begin{array}{cc} 0 & 0 \\ iJ/\hbar & -iJ/\hbar \\
-iJ/\hbar & iJ/\hbar \end{array} \right), \]
\[ {\cal C}=\left(\begin{array}{ccc} 0 & iJ/\hbar & -iJ/\hbar \\
0 & -iJ/\hbar & iJ/\hbar \end{array}\right), \]
\be {\cal D}=\left(\begin{array}{cc}
-\frac{1}{2}(\gamma_{\downarrow}+\Gamma_{\downarrow})+i(\epsilon_2-
\epsilon_1)/\hbar & 0 \\
 0 & -\frac{1}{2}(\gamma_{\downarrow}
+\Gamma_{\downarrow})-i(\epsilon_2-\epsilon_1)/\hbar \end{array}\right).
\label{blocks} \ee
From the whole set of 9 equations for elements $\rho_{ij}(t)$,
$i,\,j=0,1$ or $2$, we have in (\ref{set1}) omitted those ones
that are separated (the whole set factorizes) and are not important below.
Index $0$ corresponds to the unexcited state where there is no exciton in
the system. As for, e.g., the zero element in the left lower corner of the
square matrix, it is owing (and corresponds) to the above stressed
importance of the existence of two baths. The following notation has been
used:
\[\gamma_{\uparrow}=\frac{2\pi}{\hbar}\frac{1}{N}\sum_{\kappa}
(g_{\kappa}\hbar\omega_{\kappa})^2
\frac{1}{\exp(\beta_1\hbar\omega_{\kappa})-1}
\delta(\hbar\omega_{\kappa}-\epsilon_1), \]
\[\gamma_{\downarrow}=\frac{2\pi}{\hbar}\frac{1}{N}\sum_{\kappa}
(g_{\kappa}\hbar\omega_{\kappa})^2[1+
\frac{1}{\exp(\beta_1\hbar\omega_{\kappa})-1}]
\delta(\hbar\omega_{\kappa}-\epsilon_1), \]
\[ \Gamma_{\uparrow}=\frac{2\pi}{\hbar}\frac{1}{N}\sum_{\kappa}
(G_{\kappa}\hbar\omega_{\kappa})^2
\frac{1}{\exp(\beta_2\hbar\omega_{\kappa})-1}
\delta(\hbar\omega_{\kappa}-\epsilon_2), \]
\be \Gamma_{\downarrow}=\frac{2\pi}{\hbar}\frac{1}{N}\sum_{\kappa}
(G_{\kappa}\hbar\omega_{\kappa})^2[1+
\frac{1}{\exp(\beta_2\hbar\omega_{\kappa})-1}]
\delta(\hbar\omega_{\kappa}-\epsilon_2). \label{gamdef} \ee
Implicitly, we assume everywhere the thermodynamic limit of the bath(s) to
be already performed. As for $\beta_1$ and $\beta_2$, notice that we have
assumed the bath to consist of two (sub-)baths designated as $1$ and $2$,
each of them connected (forming inherent part thereof) with its own
subsystem I or II. In order to get rid of the initial condition, i.e.
inhomogeneous, terms (otherwise resulting in the set (\ref{set1})), we have
assumed that the bath is initially statistically
independent of the system. We impose further condition here: As the two
sub-baths do not directly interact, we ascribe both of them initial
canonical distributions with presumably different initial temperatures
$T_1$ and $T_2$. Then $\beta_j=1/(k_BT_j)$, $j=1,\ 2$. Unless the
opposite gets mentioned explicitly below, however, we shall assume the two
initial temperatures of the baths equal, i.e.
$\beta_1=\beta_2\equiv\beta=1/(k_BT)$. Then only one type of the
Bose-Einstein distribution $n_B(z)=[\exp(\beta z)-1]^{-1}$ for
phonons enters the above formulae. Worth mentioning here is
also the fact that all the $\gamma$'s and $\Gamma$'s in (\ref{gamdef})
result as standard Golden Rule transition rates (fulfilling the usual
detailed balance conditions) between states of a localized and absent
exciton. No $J$ appears in these formulae. Also this is one of
consequences of our above scaling procedure that leads to description of
relaxation not in the weak-coupling but rather intermediate or strong
coupling regimes (comparable dephasing which is due to the coupling to the
bath, and in-phasing which is owing to the above $J$-term (hopping term)).

\section{Relation to the weak-coupling limit dynamics}

The weak-coupling limit dynamics is best described in terms of eigenstates
of the Hamiltonian of the system split off the bath
\be
H_S=\sum_{j=1}^2\epsilon_ja_j^{\dagger}a_j+J(a_1^{\dagger}a_2
+a_2^{\dagger}a_1).\label{H_S} \ee
In our case, one such an eigenstate is known, it is the ground state
$|0\rangle$ of the electronic system (no exciton in the system). As for
the next two states, they read for $\epsilon_1\neq\epsilon_2$
\be |+\rangle=\chi |1\rangle-\phi |2\rangle, \;
|-\rangle=\phi |1\rangle+\chi|2\rangle \label{sta1} \ee
where
\[ \phi=\frac{2J{\rm sign}(\epsilon_2-\epsilon_1)}{\sqrt{2\sqrt{(\epsilon_2-
\epsilon_1)^2+4J^2}[|\epsilon_2-\epsilon_1|+\sqrt{(\epsilon_2-\epsilon_1)^2
+4J^2}]}}, \]
\be \chi=\sqrt{1-\phi^2}=\frac{|\epsilon_2-\epsilon_1|+\sqrt{(\epsilon_2-
\epsilon_1)^2 +4J^2}
}{\sqrt{2\sqrt{(\epsilon_2-\epsilon_1)^2+4J^2}[|\epsilon_2-\epsilon_1|
+\sqrt{(\epsilon_2-\epsilon_1)^2 +4J^2}]}}. \label{coef} \ee
The corresponding eigenenergies (unperturbed by the coupling to the bath)
read
\be E_{\pm}=\frac{1}{2}[\epsilon_1+\epsilon_2\mp{\rm sign}(\epsilon_2-
\epsilon_1)\sqrt{(\epsilon_2-\epsilon_1)^2+4J^2}]. \label{Energ} \ee
Still for the above case, $|+\rangle\rightarrow|1\rangle$ and
$|-\rangle\rightarrow|2\rangle$ when $J\rightarrow0$.
For simplicity, let us consider just the case of
$\epsilon_1\neq\epsilon_2$. In the basis of states $|0\rangle$,
$|+\rangle$, and $|-\rangle$, the rigorous weak-coupling dynamics
\cite{Davi1,Davi2} as obtained by scaling (\ref{VanHove})
of the exciton density matrix with the above initial condition reads
\[\frac{d}{dt}\left(\begin{array}{c}\rho_{00}(t) \\ \rho_{++}(t) \\
\rho_{--}(t) \\ \rho_{0+}(t) \\ \rho_{+0}(t) \\ \rho_{0-}(t) \\
\rho_{-0}(t) \\ \rho_{+-}(t) \\ \rho_{-+}(t) \end{array}\right)=
\left(\begin{array}{ccc} \ldots & \dots & \dots\\ \ldots & {\cal EN}-{\cal
DIF} & \ldots \\ \ldots & \ldots & \ldots \end{array}\right)
\cdot\left(\begin{array}{c} \rho_{00}(t) \\ \rho_{++}(t) \\
\rho_{--}(t) \\ \rho_{0+}(t) \\ \rho_{+0}(t) \\ \rho_{0-}(t) \\
\rho_{-0}(t) \\ \rho_{+-}(t) \\ \rho_{-+}(t) \end{array}\right)\]
\[+\left(\begin{array}{ccccccccc} -\gamma_{+0}-\Gamma_{-0} & \gamma_{0+} &
\Gamma_{0-} & 0 & 0 & 0 & 0 & 0 & 0 \\
\gamma_{+0} & -\gamma_{0+} & 0 & 0 & 0 & 0 & 0 & 0 & 0 \\
\Gamma_{-0} & 0 & -\Gamma_{0-} & 0 & 0 & 0 & 0 & 0 & 0 \\
0 & 0 & 0 & k & p & 0 & 0 & 0 & 0 \\
0 & 0 & 0 & p & k^* & 0 & 0 & 0 & 0 \\
0 & 0 & 0 & 0 & 0 & l & q & 0 & 0 \\
0 & 0 & 0 & 0 & 0 & q & l^* & 0 & 0 \\
0 & 0 & 0 & 0 & 0 & 0 & 0 & m & 0 \\
0 & 0 & 0 & 0 & 0 & 0 & 0 & 0 & m^* \end{array}\right)
\cdot\left(\begin{array}{c} \rho_{00}(t) \\ \rho_{++}(t) \\
\rho_{--}(t) \\ \rho_{0+}(t) \\ \rho_{+0}(t) \\ \rho_{0-}(t) \\
\rho_{-0}(t) \\ \rho_{+-}(t) \\ \rho_{-+}(t) \end{array}\right).\]
\be \label{WCD} \ee
Here
\[ \gamma_{+0}=\frac{2\pi}{\hbar}\frac{1}{N}\sum_{\kappa}(\chi^2g_{
\kappa}^2+\phi^2G_{\kappa}^2)(\hbar\omega_{\kappa})^2n_B(\hbar
\omega_{\kappa})\delta(E_+-\hbar\omega_{\kappa}), \]
\[ \gamma_{0+}=\frac{2\pi}{\hbar}\frac{1}{N}\sum_{\kappa}(\chi^2g_{
\kappa}^2+\phi^2G_{\kappa}^2)(\hbar\omega_{\kappa})^2[1+n_B(\hbar
\omega_{\kappa})]\delta(E_+-\hbar\omega_{\kappa}), \]
\[ \Gamma_{-0}=\frac{2\pi}{\hbar}\frac{1}{N}\sum_{\kappa}(\phi^2g_{
\kappa}^2+\chi^2G_{\kappa}^2)(\hbar\omega_{\kappa})^2n_B(\hbar
\omega_{\kappa})\delta(E_--\hbar\omega_{\kappa}), \]
\be \Gamma_{0-}=\frac{2\pi}{\hbar}\frac{1}{N}\sum_{\kappa}(\phi^2g_{
\kappa}^2+\chi^2G_{\kappa}^2)(\hbar\omega_{\kappa})^2[1+n_B(\hbar
\omega_{\kappa})]\delta(E_--\hbar\omega_{\kappa}),
                               \label{WCco1} \ee
are the Golden Rule transfer rates between eigenstates of the unperturbed
Hamiltonian of the system $H_S$; clearly, e.g. $\gamma_{+0}\rightarrow
\gamma_{\uparrow}$ when $J\rightarrow0$ etc. Further,
$\left(\begin{array}{ccc} \ldots & \dots & \dots\\ \ldots & {\cal EN}-{\cal
DIF} & \ldots \\ \ldots & \ldots & \ldots \end{array}\right)$
is the $9\times 9$ diagonal matrix with diagonal
elements $0$, $0$, $0$, $\frac{i}{\hbar}E_+$, $-\frac{i}{\hbar}E_+$,
$\frac{i}{\hbar}E_-$, $-\frac{i}{\hbar}E_-$, $\frac{i}{\hbar}(E_--E_+)$,
and $\frac{i}{\hbar}(E_+-E_-)$. Finally,
\[ k=-0.5(\gamma_{+0}+\gamma_{0+}+\Gamma_{-0}), \; \;
l=-0.5(\Gamma_{-0}+\Gamma_{0-}+\gamma_{+0}),\]
\be m=-0.5(\gamma_{0+}+\Gamma_{0-}),\; \;
p=0.5(\gamma_{+0}+\gamma_{0+}),\; \; q=0.5(\Gamma_{0-}+\Gamma_{-0}).
\label{WCco2} \ee
Topologically, the weak-coupling relaxation matrix (the square matrix
in the second term on the right hand side of (\ref{WCD})) resembles that
in our intermediate or strong-coupling case (\ref{set1}-\ref{blocks}).
These relaxation matrices are written down in different bases, however, and
in these bases, they have a simple, standard and easily understandable
form. Difference between
the `extended' basis $|0\rangle$, $|+\rangle$, and $|-\rangle$ and the
`localized' one $|0\rangle$, $|1\rangle$, and $|2\rangle$ is why the
free-motion terms (the first terms on the right hand side of (\ref{WCD})
and (\ref{set1})) formally differ. Coincidence of the free-motion terms is
of course complete once they are brought to the same (either localized or
extended) basis. This is, on the other hand, unlike the relaxation matrix.
If we set $J\rightarrow0$ in all the coefficients in the relaxation matrix
in the extended basis in the second term in (\ref{WCD}), we do {\em not}
reproduce the relaxation matrix of (\ref{set1}) in this basis. Instead, we
get, in this way, the relaxation matrix of (\ref{set1}) as it is written
down in (\ref{set1}), i.e. in the localized basis. The point
is that \begin{itemize} \item in the scaling (\ref{VanHove}) underlying the
weak-coupling regime, the in-phasing (which is due to $J$ that is kept
{\em constant} during the scaling) is automatically, as a consequence of
(\ref{VanHove}), presumed {\em dominating}
over dephasing processes. Nothing is changed on this feature even if we
additionally take $J$ arbitrarily small. That is why the free-motion term
is diagonal in the extended basis and the relaxation term describes
relaxation in the same
basis, i.e. that of eigenstates of $H_S$ with $J\neq0$. This relaxation
including dephasing is, owing to the form of (\ref{VanHove}), to be
understood as (in the limiting sense) infinitely slow as compared to the
in-phasing processes. On the other hand, \item with our
scaling
\be g\rightarrow0,\;J\rightarrow0,\;\tau\rightarrow+\infty,\;
g^2\tau={\rm const}, \; \frac{g^2}{J}={\rm const}, \label{VanHove2} \ee
also $J$ is scaled. Hence, as we are allowed, in such a scaling, to keep
just second order (in $g$) processes, we must set $J=0$ inside all the
relaxation superoperator (correction to ${\cal PL}(t){\cal P}\rho(t)$ on
the right hand side of (\ref{SHTS})) as the latter is already $\propto
g^2$ owing to its proportionality to the second power of the coupling to
the bath). That is why the relaxation term in (\ref{set1}) describes
relaxation to eigenstates of $H_S|_{J=0}$, i.e. in the localized basis.
$J$ must be, however, kept nonzero in the free-motion term. That is why we
get a proper competition between free-motion (no transitions in the
extended basis, i.e. between eigenstates of $H_S|_{J\neq0}$) reflecting
in-phasing owing to term $\propto J$ in (\ref{Ham}), and relaxation going
between eigenstates of $H_S|_{J=0}$, i.e. in the localized basis.
\end{itemize}
The situation in our intermediate or strong coupling case connected
with scaling (\ref{VanHove2}) thus reminds a bit of the Hubbard model when
the band and site-local interaction terms are diagonal just in the
extended and and local bases, respectively. In our case, full equivalence
between (\ref{WCD}) and (\ref{set1}) appears only in the extreme case of
$J$ taken as zero from the very beginning, in both the free-motion and
relaxation terms. This is because then the extended and localized bases
coincide.

\section{Particle and energy flows}

Let us return to our above scheme of the slow exciton dynamics as described
by (\ref{set1}-\ref{blocks}) and let us first discuss solution for the case
of zero $J$ (for this particular case, correspondence with the weak-coupling
approach as above can be found quite easily). It is easy to verify that the
stationary (and in this particular case definitely also equilibrium) solution
to (\ref{set1}) reads
\[ \rho_{00}\equiv\rho_{00}(t\rightarrow+\infty)=1-\rho_{11}-\rho_{22},
\] \be
\rho_{11}=\frac{\exp(-\beta\epsilon_1)}{Z},\;
\rho_{22}=\frac{\exp(-\beta\epsilon_2)}{Z},\; Z=1+\exp(-\beta\epsilon_1)+
\exp(-\beta\epsilon_2). \label{eqso} \ee
As for stationary values of site off-diagonal elements of $\rho$,
they turn to zero.
So as to derive (\ref{eqso}), we have used the detailed balance relations
$\gamma_{\uparrow}/\gamma_{\downarrow}=\exp(-\beta\epsilon_1)$ and
$\Gamma_{\uparrow}/\Gamma_{\downarrow}=\exp(-\beta\epsilon_2)$. Formulae
(\ref{eqso}) are still in full agreement with the equilibrium statistical
mechanics, in particular the canonical distribution. In order to see that,
let us realize that we have omitted, for purely technical reasons, the
two-exciton state with both levels ($1$ and $2$) occupied by excitons. This
means errors of the order $\propto \exp(-\beta(\epsilon_1+\epsilon_2))$.
Within this accuracy, we can well approximate, e.g., the
stationary value $\rho_{11}$ as $\rho_{11}\approx \frac{\exp(-
\beta\epsilon_1)}{1+\exp(-\beta\epsilon_1)}$ what is the canonical
equilibrium probability $P_1$ of finding the site 1 occupied by the
exciton, as prescribed by the equilibrium statistical mechanics.
Refraining, on the other hand, for a while from the above omission of the
two-exciton state and designating the two-exciton state as state $3$, we
might reconsider the problem on the more general level. This would yield
the stationary value $\rho_{11}= \exp(-\beta\epsilon_1)/Z'$ and
$\rho_{33}=\exp(-\beta(\epsilon_1+\epsilon_2))/Z'$ where $Z'=Z+
\exp(-\beta(\epsilon_1+\epsilon_2))$. From that, the probability of finding
the exciton at site $1$ irrespective of the occupation of site $2$ results
as $P_1\equiv\rho_{11}+\rho_{33}=\frac{1}{1+
\exp(\beta\epsilon_1)}$~~~~\footnote{This is the standard Fermi-Dirac
distribution for excitons that behave as paulions, i.e on-site fermions,
with zero value of their chemical potential.},
in a full correspondence with the above reasoning based on discussion of
accuracy of our treatment.

Interesting and important for what follows below is also the dynamics of
occupation of site $1$. As we still keep $J=0$, we may for this
purpose ignore the state $2$ at all. Doing so, we shall for a
while completely neglect
$\gamma_{\uparrow}$ as compared to $\gamma_{\downarrow}$. \footnote{It is
always $\gamma_{\uparrow}/\gamma_{\downarrow}=\exp(-\beta\epsilon_1)$ (the
detailed balance condition). This ratio is definitely $\ll 1$ for
$k_BT\ll\epsilon_1$. On the other hand, except in (\ref{exde}), this
condition is not used below.} Then the dynamics
(time dependence) of probability of finding the exciton at site $1$ reads
\be P_1(t)=P_1(t=0){\rm e}^{-\gamma_{\downarrow}t}. \label{exde} \ee
This corresponds to the probability amplitude of finding the exciton at
site $1$ in form of ${\rm e}^{-i\epsilon t/\hbar-\gamma_{\downarrow}t
/2}$ whose Fourier transform reads as Lorentzian $\frac{1}{\pi}\frac{
\gamma_{\downarrow}/2}{[\omega-\epsilon_1/\hbar]^2+
[\gamma_{\downarrow}/2]^2}$. This indicates that the exciton level
$\epsilon_1$ (and similarly for the level $\epsilon_2$) is broadened as a
Lorentzian with the half-width $\gamma_{\downarrow}/2$. This is what we
shall need below.

All this is very reasonable and known, in different context, for
already many years. In what follows, we return to a general situation. We
shall argue now that the matter will drastically change once we put $J$
nonzero. To be more concrete, we are going to argue that under the above
defined conditions and for still equal initial temperatures of the
baths $T_1=T_2=T$, there will be a permanent exciton (and also energy)
flow between sites 1 and 2. Because of our joining these sites
with different and mutually non-interacting baths, this implies the
existence of the exciton (energy) flow also between our subsystems I and
II.

This prediction is perhaps shocking for standardly thinking physicists. In
order to deprive potential opponents of tempting, at the first sight fully
legitimate but still insufficiently justified arguments, let us already
here state (admit) two important things right here: \begin{itemize} \item
The weak-coupling theory does not yield the persistent nonzero flows.
\item Application of the canonical distribution to the whole system
(consisting of two subsystems, each of them having its own bath and
electronic, i.e., exciton levels) also yields that these flows are in
average zero. \end{itemize}
The counter-arguments against such an easy beating off our type of
reasoning and results are, however, that \begin{itemize} \item The
weak coupling theory has even potentially no possibility to yield such
flows at all. So, it cannot serve as an arbitrator. The point is that such
flows, as shown below, need a sufficiently strong dephasing (as compared
with in-phasing processes) to broaden the energy levels $\epsilon_1$ and
$\epsilon_2$ (as we shall see below, the transfer is between tails of
these levels). The very definition of the weak-coupling approach (see
(\ref{VanHove}) above) is based on the Van Hove limit leading to a
negligible role of the dephasing (as compared to the in-phasing), i.e. the
approach is unable to model the situation we speak about, where
we predict the existence (nonzero values) of the flows.\footnote{Notice
also that, for, e.g., negligible relativistic corrections and in absence of
external magnetic field, the canonical density matrix is real in our local
basis $|1\rangle$ and $|2\rangle$. Hence, formula (\ref{Ide}) below yields
zero flow between sites $1$ and $2$ in the canonical equilibrium,
irrespective of the Golden-Rule-type prediction (\ref{Idef}).} Under the
condition of the dominating in-phasing (over the dephasing as in the
weak-coupling regime), exciton at sites 1 and 2 becomes shared. In
other words, a special type of a covalent bonding appears between the
sites which (as also found in other situations) prevents such flows.
\item If we are really right in our prediction that, in the
thermodynamic limit, there is a persistent flow between
sites 1 and 2, i.e. between systems I and II, application of the canonical
distribution is unjustified. The point is that the canonical distribution
is based on maximizing entropy under solely two constraints: Normalization
condition ($Tr \rho=1$) and mean energy conservation ($Tr (H\rho)=\bar{E}
=$const). If we are right, then at least the third constraint $Tr
(\hat{I}\rho)=$const ($\hat{I}$ being the flow operator) should be added
what makes the usual canonical distribution improper. Also other
approaches used to justify the canonical distribution always use,
explicitly or implicitly, the {\em ad hoc} assumption of non-existence of
other persistent quantities than energy. So, if we are really right in our
arguments here, then, definitely, nor the canonical distribution can be
used as an arbiter.
\end{itemize}

Let us start our reasoning here by deriving a sufficiently reliable form
of the exciton flow formula. From our Hamiltonian (\ref{Ham}) and the
Liouville equation, we get that
\[ -\frac{d}{dt}\langle a_1^{\dagger}a_1\rangle=\frac{i}{\hbar}
\langle[a_1^{\dagger}a_1,H]\rangle\] \be =\frac{2J}{\hbar}\Im{\rm
m}\langle a_2^{\dagger}a_1\rangle+\frac{2}{\sqrt{N}}\sum_{\kappa}
g_{\kappa}\omega_{\kappa}\Im{\rm m}\langle b_{1\kappa}^{\dagger}a_1
\rangle. \label{I1} \ee
The last mean value can be calculated, using principle of the adiabatic
switching on the interactions, as
\be \delta\langle b_{1\kappa}^{\dagger}a_1\rangle=\frac{d}{dt}\langle
b_{1\kappa}^{\dagger}a_1\rangle=-\frac{i}{\hbar}\langle [b_{1\kappa}^{
\dagger}a_1,H] \rangle,\;\delta\rightarrow0+. \label{I2} \ee
Here, terms containing $J$ should already be omitted if we work to the
second order in our small parameter $g$ only (remember that $J\propto
g^2$ - see (\ref{VanHove2})). Thus, within this accuracy,
\be \langle b_{1\kappa}^{\dagger}a_1\rangle
\approx\frac{1}{\hbar\omega_{\kappa}-
\epsilon_1+i\hbar\delta}\frac{1}{\sqrt{N}}g_{\kappa}\hbar\omega_{\kappa}
\{n_B(\hbar\omega_{\kappa})[1-\langle
a_1^{\dagger}a_1\rangle-[1+n_B(\hbar\omega_{\kappa})]\langle
a_1^{\dagger}a_1\rangle\}. \label{I3} \ee
So, (\ref{I1}) reads, within the required accuracy, as
\be -\frac{d}{dt}\langle a_1^{\dagger}a_1\rangle\approx\frac{2J}{\hbar}
\Im{\rm m}\rho_{12}-\rho_{00}\gamma_{\uparrow}+\rho_{11}
\gamma_{\downarrow}. \label{I4} \ee
As the last two terms express the exciton number imbalancing owing to
transfers $1\leftrightarrow0$, the proper formula for the real
$1\leftrightarrow2$ flow is connected with the first term on the right hand
side of (\ref{I4}). Thus, the $1\leftrightarrow2$ exciton flow (taken as
positive if flowing from 1 to 2) reads
\be I=\frac{2J}{\hbar}\Im{\rm m}\rho_{12}. \label{Ide} \ee
The fact that $I$ is determined by the (imaginary part of the) site
off-diagonal elements of the particle density matrix follows already from
the elementary quantum mechanics.

The long-time (stationary) value of the $\rho_{12}$ element of the density
matrix can be found, however, from (\ref{set1}) (by setting the
time-derivatives zero), incorporating also the normalization condition
\be \sum_{j=0}^2\rho_{jj}=1. \label{NorCon} \ee
After a simple algebra, the result is
\[ I=\frac{2\pi}{\hbar}J^2\cdot\frac{1}{\pi}\frac{\frac{\hbar}{2}(\gamma_{
\downarrow}+\Gamma_{\downarrow})}{[\frac{\hbar}{2}(\gamma_{\downarrow}+
\Gamma_{\downarrow})]^2+[\epsilon_2-\epsilon_1]^2}\] \be\cdot
\frac{\gamma_{\uparrow}\Gamma_{\downarrow}-\gamma_{\downarrow}\Gamma_{
\uparrow}}{\gamma_{\downarrow}\Gamma_{\downarrow}+\gamma_{\downarrow}\Gamma_{
\uparrow}+\gamma_{\uparrow}\Gamma_{\downarrow}+X[\gamma_{\downarrow}+
\Gamma_{\downarrow}+2\gamma_{\uparrow}+2\Gamma_{\uparrow}]} \neq 0.
\label{Iex1} \ee
Here
\be X=\frac{2\pi}{\hbar}J^2\cdot\frac{1}{\pi}\frac{\frac{\hbar}{2}(
\gamma_{\downarrow}+\Gamma{\downarrow})}{[\frac{\hbar}{2}(
\gamma_{\downarrow}+\Gamma_{\downarrow})]^2+[\epsilon_2-\epsilon_1]^2}.
\label{Xdef} \ee

Since we are obliged to stick to the required accuracy, we can deal, using
the formalism allowed, with just the leading terms. Hence, we shall omit
the terms $\propto X$ in the denominator assuming that
\be X\ll\frac{\gamma_{\downarrow}\Gamma_{\downarrow}}{\gamma_{\downarrow}
+\Gamma_{\downarrow}}. \label{SmX} \ee
This means that, up to terms of higher than sixth order in $g$ (still
remember that $J\propto g^2$),
\[ I\approx\frac{2\pi}{\hbar}J^2\cdot\frac{1}{\pi}\frac{\frac{\hbar}{2}(
\gamma_{\downarrow}+\Gamma_{\downarrow})}{[\frac{\hbar}{2}(\gamma_{
\downarrow}+\Gamma_{\downarrow})]^2+[\epsilon_2-\epsilon_1]^2}\] \[
\cdot\frac{\gamma_{\uparrow}\Gamma_{\downarrow}-\gamma_{\downarrow}\Gamma_{
\uparrow}}{\gamma_{\downarrow}\Gamma_{\downarrow}+\gamma_{\downarrow}\Gamma_{
\uparrow}+\gamma_{\uparrow}\Gamma_{\downarrow}} \]
\be =\frac{2\pi}{\hbar}J^2\cdot\frac{1}{\pi}\frac{\frac{\hbar}{2}(\gamma_{
\downarrow}+\Gamma_{\downarrow})}{[\frac{\hbar}{2}(\gamma_{\downarrow}
+\Gamma_{\downarrow})]^2+[\epsilon_2-\epsilon_1]^2}[\rho_{11}-\rho_{22}].
\label{Idef} \ee
Interesting point is that the expression for the exciton flow on the right
hand side of (\ref{Idef}) is correct even without assuming (\ref{SmX}).
This follows from (\ref{Iex1}) by taking into account that from
(\ref{set1}), we obtain the asymptotic-time populations
\[ \rho_{11}=\frac{\gamma_{\uparrow}\Gamma_{\downarrow}+
X(\gamma_{\uparrow}+\Gamma_{\uparrow})}{\gamma_{\downarrow}\Gamma_{
\downarrow}+\gamma_{\downarrow}\Gamma_{\uparrow}+\gamma_{\uparrow}
\Gamma_{\downarrow}+X(\gamma_{\downarrow}+\Gamma_{\downarrow}+
2\gamma_{\uparrow}+2\Gamma_{\uparrow})}\approx\frac{{\rm
e}^{-\beta\epsilon_1}}{Z}, \]
\be \rho_{22}=\frac{\gamma_{\downarrow}\Gamma_{\uparrow}+
X(\gamma_{\uparrow}+\Gamma_{\uparrow})}{\gamma_{\downarrow}\Gamma_{
\downarrow}+\gamma_{\downarrow}\Gamma_{\uparrow}+\gamma_{\uparrow}
\Gamma_{\downarrow}+X(\gamma_{\downarrow}+\Gamma_{\downarrow}+
2\gamma_{\uparrow}+2\Gamma_{\uparrow})}\approx\frac{{\rm
e}^{-\beta\epsilon_2}}{Z}. \label{Popul} \ee
In the last approximate expressions, we have again used condition
(\ref{SmX}).

Clearly, expression (\ref{Idef}) for the exciton flow is, quite
surprisingly at the first sight, clearly nonzero. Already this is
remarkable as we have to realize again that the exciton transfers
energy and the transfer channel 1$\leftrightarrow$2 is the only channel
connecting our systems I and II and able, within our model, to transfer
energy between them. Before getting into more physical details connected
with this observation, let us also comment that expression (\ref{Idef})
\begin{itemize} \item has a proper total balance structure with transitions
$1\rightarrow2$ and $2\rightarrow1$ (contributing to (\ref{Idef}) by
terms
$\propto\rho_{11}$ and $\propto\rho_{22}$, respectively), and \item is
fully compatible with (in fact, it is exactly equal to) the second-order
Golden Rule of quantum mechanics ($J$ is the matrix element of the
transfer part of the Hamiltonian between states of the exciton at sites
$1$ and $2$) assuming that the energy conservation law is properly
broadened owing to exciton decay processes. \end{itemize} This broadening
means that the exciton transfer is neither at level
$\epsilon_1$ nor at level $\epsilon_2(\neq\epsilon_1)$ but generally at
arbitrary energy in tails of the two broadened exciton levels. (Realize
that the exciton is, as generally in nature and as also anticipated in our
model, just a finite life-time quasiparticle.) This interpretation is
clearly confirmed by the fact that one can rewrite (\ref{Idef}) also as
\be I=\frac{2\pi}{\hbar}J^2[\rho_{11}-\rho_{22}]\cdot
\int_{-\infty}^{+\infty}\frac{1}{\pi}\frac{\frac{\hbar}{2}
\gamma_{\downarrow}}{[\frac{\hbar}{2}
\gamma_{\downarrow}]^2+[\epsilon-\epsilon_1]^2}\cdot\frac{1}{\pi}
\frac{\frac{\hbar}{2}\Gamma_{\downarrow}}{[\frac{\hbar}{2}\Gamma_{
\downarrow}]^2+[\epsilon-\epsilon_2]^2}\,d\epsilon \label{Idef1} \ee
and the fact that any quasiparticle exponentially damped with the decay
rate $\gamma$, i.e. the survival probability amplitude
\be \langle a_1(t)a_1^{\dagger}\rangle=
\exp(-i\epsilon_1\cdot t/\hbar-\gamma t/2), \label{quasi} \ee
has its energy level (here $\epsilon_1$) broadened into a Lorentzian with
the energy half-with $\hbar\gamma/2$. See also a comment in this respect
above. The forms of (\ref{Idef}) and (\ref{Idef1}) thus leave only very
limited space for speculations about validity of our approach.

Our statement about persistent energy transfer between subsystems I and II
kept at (initially) equal temperatures is already justified by our
expressions for the exciton flow $I$. Nevertheless, let us raise the (in a
way) subsidiary question what is the energy flow between the two
subsystems. According to the above quasiparticle interpretation, one would
expect that the energy flow is
\[ Q= \frac{2\pi}{\hbar}J^2[\rho_{11}-\rho_{22}]\cdot
\int_{-\infty}^{+\infty}\frac{1}{\pi}\frac{\frac{\hbar}{2}
\gamma_{\downarrow}}{[\frac{\hbar}{2}
\gamma_{\downarrow}]^2+[\epsilon-\epsilon_1]^2}\cdot\frac{1}{\pi}
\frac{\frac{\hbar}{2}\Gamma_{\downarrow}}{[\frac{\hbar}{2}\Gamma_{
\downarrow}]^2+[\epsilon-\epsilon_2]^2}\,\epsilon \,d\epsilon \]
\be =\frac{2\pi}{\hbar}J^2[\rho_{11}-\rho_{22}]\cdot\frac{\hbar}{2\pi}
\frac{\epsilon_2\gamma_{\downarrow}+\epsilon_1\Gamma_{\downarrow}}{
[\frac{\hbar}{2}(\gamma_{\downarrow}+\Gamma_{\downarrow})]^2
+[\epsilon_2-\epsilon_1]^2}. \label{Qdef} \ee
The problem is with general justification of this formula. In fact, one
should define the energy flow $Q$ between systems I and II in full
generality as
\be Q=-\frac{d}{dt}\langle \bar{H}_{I}\rangle \label{Ene1} \ee
or equivalently
\be Q=\frac{d}{dt}\langle \bar{H}_{II}\rangle. \label{Ene2} \ee
Here $\bar{H}_{I}$ and $\bar{H}_{II}$ should have the meaning of energies
of the subsystems I and II, such, that $\bar{H}_{I}+\bar{H}_{II}=H$ (in
order to have (\ref{Ene1}) compatible with (\ref{Ene2})). Because $J$ must
be assumed nonzero, these evidently cannot be the Hamiltonians $H_{I}$
and $H_{II}$ introduced in (\ref{Ham}). Really, making this or any other
trivial identification of $\bar{H}_{I}$ and $\bar{H}_{II}$ leads to
hardly interpretable results. The physical reason for that is that nonzero
values of $J$ cause effects like $J$-dependent renormalization of the
exciton coupling to the bath. Its exact form is unknown so that one can
ignore it only when the corresponding coupling constants are negligibly
small (when there is nothing to be renormalized). That is why, for very
small $g_{\kappa}$ (i.e. negligible $\gamma_{\downarrow}$),
one can define the energy flow properly and reliably by (\ref{Ene1}),
identifying (in this particular case) $\bar{H}_{I}$ with $H_{I}$. Then
(\ref{Ene1}) reduces to the above formula (\ref{Qdef}) with negligible
$\gamma_{\downarrow}$. In the opposite limiting case, when $G_{\kappa}$ gets
very small (i.e. with negligible $\Gamma_{\downarrow}$),
(\ref{Ene2}) also reduces to the above formula (\ref{Qdef}), this time
with negligible $\Gamma_{\downarrow}$, provided we identify
$\bar{H}_{II}$ with $H_{II}$.

So summarizing: \begin{itemize} \item Suggested interpolation formula for
the energy flow (\ref{Qdef}) can be properly justified in the two above
mentioned limiting cases. Then it definitely yields nonzero values of the
energy flow between our subsystems when $\epsilon_1\neq\epsilon_2$ and
still $T_1=T_2$ (equal initial temperatures of the two baths). \item
Formula for the exciton flow (\ref{Idef}), or its equivalent form
(\ref{Idef1}), can be, in the above way, properly justified for all the
values of the parameters involved. It always, for
$\epsilon_1\neq\epsilon_2$ and $T_1=T_2$, yields nonzero exciton flow
between the subsystems, always in the direction from the system with
higher population to that one with the lower population of the
corresponding exciton level. As the excitons bear energy, this clearly
implies (and fully corresponds to the above) {\em nonzero} mean energy
flow.\end{itemize}

The last point to be stressed here results from rather a trivial reasoning
based on the above formulae and continuity, with respect to, e.g.,
temperature $T_2$, of the results obtained. On the other hand, it is
of highest importance from the point of view of interpretation: It is
easy to see that if we take (as
always here) energies of the exciton levels in our two systems different,
i.e. $\epsilon_1\ne\epsilon_2$, but allow the initial temperatures of
the two subsystems be different ($T_1\ne T_2$), we can, for any values of
$\epsilon_1\ne\epsilon_2$ and arbitrary $T_1$, find such a value of the
initial temperature $T_2$ of the second subsystems that
`separate-equilibrium' (for $J=0$, i.e. for separated subsystems I and II)
values of population of the disconnected exciton levels $\epsilon_1$ and
$\epsilon_2$ become equal. In other words, $\rho_{11}=\rho_{22}$. According
to what has been argued above, however, this makes, upon reintroducing the
(sufficiently weak - see (\ref{SmX})) exciton transfer channel (setting
$J$ nonzero though small enough in the sense of (\ref{SmX})) both the
exciton $I$ and the energy flows $Q$ (simultaneously) zero -  see
formulae (\ref{Idef}), (\ref{Idef1}), and (\ref{Qdef}).

In particular, assume now that, e.g., $\epsilon_2>\epsilon_1$. Then
clearly, arguing just from the above `separate-equilibrium' values of
$\rho_{11}$ and $\rho_{22}$, the temperature $T_2$ for which the values
of $\rho_{11}$ and $\rho_{22}$ get equal, is greater that $T_1$. Let us
call this temperature $T_2^{crit}$. It is easy to show that $T_2^{crit}=
T_1\epsilon_2/\epsilon_1$. Clearly, for temperatures
$T_2$ greater than $T_1$ but less than $T_2^{crit}$, the energy (exciton)
flow is nonzero and still going in the direction from the subsystem I
with the higher exciton population (lower temperature $T_1<T_2$ but
essentially less excitation energy $\epsilon_1<\epsilon_2$) to that one
with the lower exciton concentration - subsystem II. This means, the
{\em exciton and energy flow goes}, fully surprisingly, {\em against the
temperature step}. This conclusion is of highest importance.

\section{Towards the second law of thermodynamics}

There are several formulations of the second law. The form by Clausius
from 1865 involves entropy that was not discussed here. Existence of
entropy is, however, in fact consequence of three main formulations of the
second law, that one by Thomson (1849), Clausius (1850), and
Carath\'{e}odory (1909) (for connections to
entropy see \cite{LieYng} or standard textbooks on axiomatic
thermodynamics). The statements are (cited according to \cite{LieYng}):
\begin{description}
\item[Thomson] (as Baron Kelvin of Largs since 1892) \cite{Thom}:
{\em No process is possible, the sole result of which is that a body is
cooled and work is done.}
\item[Clausius] \cite{Clau}: {\em No process is possible the sole result
of which is that the heat is transferred from a body to a hotter one.}
\item[Carath\'{e}odory] \cite{Cara}: {\em In any neighbourhood of any state
there are states that cannot be reached from it by an adiabatic process.}
\end{description}
The words `sole' imply in particular that \begin{itemize} \item in the
Thomson formulation, the process should be cyclic, without any
compensation (additional heat transfer to another and cooler body).
A (thought) machine working in such a style is often called `perpetuum
mobile of the second kind'; \item in the Clausius formulation, the process
is not aided from outside. \end{itemize}

Significance of this law and consequences of its potential violation
were perhaps best described in the Introduction of \cite{LieYng}. For
150 years, nobody really questioned this statement based on uncountable
number of observations from our everyday life. One should add and stress
therefore that our thought (i.e. still not real in the sense of
in-nature-existing) systems governed by quantum mechanics and comprising
macroscopic baths are macroscopic, i.e. they should obey the Clausius (and
not only this) formulation of the second law {\em provided} that the
quantum mechanics and thermodynamics are always, including the quantum-
and at least the macro-world, compatible. Sometimes, even absolutistic
statements in favour of the unconditional validity the second law in the
macroworld stemming from our everyday experience appear ("...No
exception to the second law of thermodynamics has ever been found - not
even a tiny one..." \cite{LieYng2}). As our conclusion at the
end of the previous section shows, the opposite is true in our model here.
One should mention here the `pawl and ratchet' systems originally suggested
by Feynman \cite{Feyn} which are also often cited also in connection with
the second law. These systems, however, so far fail in practical attempts
to violate the second law \cite{Muss}, in full agreement with the Feynman
\cite{Feyn} analysis. On the contrary, behaviour of the above model does,
as argued above, contradict the second law. It is of course not for the
first time that such a mathematically well-founded behaviour incompatible
with standard thermodynamics appears - see \cite{CapBok2,CapMan} and, for
other models, papers cited above or therein. The present model is, on the
other hand, perhaps the simplest one. Worth mentioning is also that the
above criteria for the energy flow against the temperature step may easily
be compatible with, e.g., even room or higher temperatures. This is
unlike, e.g., \cite{AllNie}. In the infinite temperature (i.e. in the
classical) limit, however, the effect disappears.

\section{Towards the zeroth law of thermodynamics}

In order to be specific, let us state what this law (so often, especially
in the mechanical context, understood as trivial) says: {\em If system A is
in equilibrium with systems B and C then B is in equilibrium with C} (see,
e.g., \cite{PliBer}). It helps to introduce thermodynamic temperature,
chemical potential etc. Though {\em universal} validity of this law has
already been questioned (as far as its form for equality of chemical
potential of one sort of species in different phases in equilibrium is
concerned) - see \cite{CaSimem} or in the implicit form in \cite{Cap2}, no
special attention has so far been paid to this fact. That is why we should
address the question, in connection with our model above, again.

Let us fix, in the above model, again the situation with
$\epsilon_1<\epsilon _2$ and be $T_1$ arbitrary positive. Then clearly
$T_2^{crit}\equiv T_1\epsilon_2/\epsilon_1>T_1$ and we set the initial
temperature of the second bath $T_2=T_2^{crit}$. Hence, upon establishing
a contact between subsystems I and II by taking $J$ slightly (in the sense
of (\ref{SmX})) nonzero, we get no energy or exciton flow. Now, we can
invoke standard thermodynamic definition of what it means to say that
two bodies in a contact are in mutual equilibrium. The definition reads
that introducing arbitrary obstacles hindering flows between the bodies
does not change their thermodynamic state. In this sense, this is exactly
the situation we are now in: We have two bodies in a contact where there
are no flows between them. Hence, the thermodynamic state cannot be
violated by any obstacle setting these flows zero and not influencing
otherwise the state of the systems because the flows are already zero.
(Notice that no other flows but those of exciton or energy can exist in
model.) Remind, on the other hand, that we have such a strange
thermodynamic equilibrium (in the above thermodynamic sense) that
temperatures of both the systems (those of their baths) are different.
This can clearly lead to other contradictions with the standard
thermodynamics as we are now going to show.

Assume now that we have still another (third) exciton level
and still another (third) thermal bath attached to it. In other words, we
complement our Hamiltonian by terms
\[ \Delta H=\epsilon_3a_3^{\dagger}a_3
+\sum_{\kappa}\hbar\omega_{\kappa}b_{3\kappa}^{\dagger}b_{3\kappa}
+\frac{1}{\sqrt{N}}\sum_{\kappa}h_{\kappa}\hbar\omega_{\kappa}(a_3+a_3
^{\dagger})(b_{3\kappa}+b_{3\kappa}^{\dagger})\]
\be +\frac{1}{\sqrt{N}}\sum_{\kappa}H_{\kappa}\hbar\omega_{\kappa}
(a_2^{\dagger}a_3+a_3^{\dagger}a_2)(b_{2\kappa}+b_{3\kappa}+
b_{2\kappa}^{\dagger}+b_{3\kappa}^{\dagger})
+K(a_3^{\dagger}a_1+a_1^{\dagger}a_3). \label{adHam} \ee
Clearly, the last two terms on the right hand side of (\ref{adHam}) provide
interaction of subsystems II and III (this type of the phonon-assisted
interaction exists just in the diabatic (non-rigid) basis \cite{SilCap})
and that of the subsystems III and I (the latter interaction and the
induced exciton transfer is for simplicity assumed coherent, like that one
of the subsystems I and II). Assume also that the exciton energy
$\epsilon_3$ equals to that of the exciton in subsystem I, i.e.
$\epsilon_3=\epsilon_1$. Next, we assume that again, the initial density
matrix is factorizable into a product of density matrices of all the
subsystems, so that there are no exciton-bath initial statistical
correlations between any two of the three subsystems. Finally, assume that
the density matrices of all the baths are initially canonical.
Corresponding temperatures are assumed as $T_3=T_2>T_1$ (the last
inequality being assumed already above).

Let us for a while set $J=K=0$. Then we have our subsystem I fully
separated and the dynamics goes between subsystems II and III only. Exactly
in the same way as above (i.e. using the same type of scaling), we get a
closed set of equations for the matrix elements of the exciton system. This
time, however, the situation is simpler as compared to that above. First,
our coupling between subsystems II and III is assumed as bath-assisted. In
connection with that, the set of equations for the site diagonal as well
as site off-diagonal matrix elements of the exciton density matrix
factorizes so that equations comprising the diagonal elements contain
{\em only the diagonal} elements (that get separated from the set for the
off-diagonal elements), reducing in form to the Pauli master equations
\cite{Paul}. Properties of these equations are sufficiently known.  So we
shall not repeat the calculations and refer the interested reader to any
elementary textbook of kinetic theory. The result for the asymptotic
exciton occupation probabilities reads as in the standard equilibrium
statistical thermodynamics, i.e.
\[\rho_{22}=\frac{\exp(-\beta_2\epsilon_2)}{1+\exp(-\beta_2\epsilon_2)
+\exp(-\beta_2\epsilon_1)}, \]
\be \rho_{33}=\frac{\exp(-\beta_2\epsilon_1)}{1+\exp(-\beta_2\epsilon_2)
+\exp(-\beta_2\epsilon_1)},\; \beta_2=\frac{1}{k_BT_2}. \label{pop2} \ee
Let us stress the following points: \begin{itemize} \item
Values (\ref{pop2}) properly reproduce, within our approximation
(omission of multi-exciton states, i.e. with errors $\propto\exp(
-\beta_2(\epsilon_2+\epsilon_1))$), standard equilibrium statistical
mean exciton numbers at sites $2$ and $3$,
i.e. $1/[\exp(\beta_2\epsilon_2)+1]$ and $1/[\exp(\beta_2\epsilon_1)+1]$
(see a comment above concerning appearance of these Fermi-Dirac
distributions for excitons). In fact, as already argued above,
reintroducing the multiple-exciton states would reproduce these values
exactly. These Fermi-Dirac distributions are, on the other hand, proper
mean number of excitons at sites $2$ and $3$ for $H_{\kappa}=0$, i.e.
separated subsystems II and III. Hence, establishing or cancelling the
above contact between the latter two subsystems does not change the
stationary (equilibrium) exciton populations at the corresponding sites.
The same may be shown to apply to phonon populations in the corresponding
baths. \item Assume now $H_{\kappa}\neq0$. Owing to the incoherent
(bath-assisted) character of the above coupling between subsystems II and
III, different populations of levels 2 and 3 (in accordance with standard
statistical thermodynamics) do not contradict the fact that (exciton
mediated) flows between subsystems II and III remain in equilibrium
exactly zero. This is due to the fact that real
$2\rightarrow3$ and $3\rightarrow2$ transition rates are proportional to
$\rho_{22}\times\{1+1/[\exp(\beta_2\hbar\omega_{\kappa})-1]\}$ and
$\rho_{33}\times1/[\exp(\beta_2\hbar\omega_{\kappa})-1]\}$, respectively.
Here, by the energy conservation law, $\hbar\omega_{\kappa}=\epsilon_2
-\epsilon_1>0$. The multiplicative factors at $\rho_{22}$ and $\rho_{33}$
are phonon statistical factors describing phonon-assisted induced as well
as spontaneous processes. So, the transfer rates $2\rightarrow3$ and
$3\rightarrow2$ are in equilibrium exactly equal, mutually cancelling their
contribution to the exciton as well as energy flow between subsystems II
and III (transferred energy is $\epsilon_1+\hbar\omega_{\kappa}=\epsilon_2
$). Here, for simplicity, we have assumed
(in accordance with assumptions underlying validity of the Pauli equations)
that $|\epsilon_2-\epsilon_3|/\hbar\equiv|\epsilon_2-\epsilon_1|/\hbar$ is
appreciably greater than the sum of broadenings of levels $2$ and $3$, i.e.
that the transitions are practically (exciton+phonon) energy conserving.
These arguments are what underlies the detailed balance conditions in the
Pauli master equation theories yielding the same conclusion.
\end{itemize}

These are the characteristics of the mutual equilibrium (according to
the above thermodynamic definition) state of subsystems II and III, and
also of the internal equilibrium states of the isolated subsystems II and
III taken separately. With this states, let us now put $H_{\kappa}=0$
(we split subsystems II and III), keep $J=0$ but put $K\neq 0$. Let us
repeat: We have the two subsystems (I and III) with equal exciton energies
$\epsilon_1=\epsilon_3$ (this case may be treated as a limit
$\epsilon_1-\epsilon_3\rightarrow0$ in the above formulae), initially in
internally canonical states of both
the subsystems (and their baths), with the respective temperatures $T_1
<T_3$. In accordance with what has been said above about development
of two such subsystems (subsystems I and II above) with their respective
baths, the asymptotic (stationary) populations of the exciton levels 1 and
3 only slightly change upon establishing contact between the subsystems.
Asymptotically, they read
\[ \rho_{11}=\frac{\exp(-\beta_1\epsilon_1)}{1+\exp(-\beta_1\epsilon_1)
+\exp(-\beta_2\epsilon_1)}\approx \exp(-\beta_1\epsilon_1),\]
\be \rho_{33}=\frac{\exp(-\beta_2\epsilon_1)}{1+\exp(-\beta_1\epsilon_1)
+\exp(-\beta_2\epsilon_1)}\approx \exp(-\beta_2\epsilon_1),\;
\beta_1=\frac{1}{k_BT_1}.\label{pop3} \ee
Clearly, because $\beta_2<\beta_1$, the populations $\rho_{11}$ and
$\rho_{22}$ are different. So, according to (\ref{Idef}), (\ref{Idef1}) as
well as (\ref{Qdef}), there are flows between the subsystems I and III,
i.e. we have no equilibrium in the thermodynamic sense.

Thus, summarizing, we have subsystems I, II and III which all have well
defined temperatures. Upon establishing just the above specific contact
between I and II, the systems stay, in the sense of the thermodynamic
definition, in equilibrium. Similarly, upon establishing just the above
contact between subsystems II and III, the thermodynamic equilibrium is
not violated. Thus, the zeroth law of thermodynamics should be, according
to thermodynamics, well applicable. It states that establishing contact
between subsystems I and III should preserve their mutual equilibrium
state. As seen above, however, {\em the opposite is true}.

\section{Conclusions}

We have investigated one standard and rather trivial model that
allows rigorous treatment by methods of the quantum theory of open systems.
The obtained behaviour contradicts what is prescribed by the second as well
as zeroth laws of thermodynamics. In connection with previously expressed
doubts about universal validity of the second law in specific situations,
this extends challenges to general compatibility of such two basic
scientific disciplines as the thermodynamics and the quantum theory.
\vspace{1cm}

{\bf Acknowledgement}

The author is deeply indebted to the Max-Planck-Institute in Stuttgart, the
University of Stuttgart, and in particular to Prof. Max Wagner for their
invitation and very kind hospitality during the author stay in Stuttgart
in September 1999. During this visit, basic ideas of the present work were
formulated and the above model was constructed. The author is also deeply
indebted to MUDr. M. Ryska (IKEM, Prague) and MUDr. A. Ivan\v{c}o (NNF,
Prague) for saving his life at the beginning of 2000 before this work was
completed. Support of grant 202/99/0182 of the Czech grant agency
is also gratefully acknowledged.  \end{document}